\documentclass{article}
\usepackage[preprint]{neurips_2023}

\usepackage[utf8]{inputenc} % allow utf-8 input
\usepackage[T1]{fontenc}    % use 8-bit T1 fonts
\usepackage{hyperref}       % hyperlinks
\usepackage{url}            % simple URL typesetting
\usepackage{booktabs}       % professional-quality tables
\usepackage{amsfonts}       % blackboard math symbols
\usepackage{nicefrac}       % compact symbols for 1/2, etc.
\usepackage{microtype}      % microtypography
\usepackage{xcolor}         % colors
\usepackage{graphicx}
\usepackage{tabularx}  % 引用 tabularx 宏包
\usepackage{booktabs}  % 用于更好的表格线条
\usepackage{array}     % 提供额外的表格功能
\usepackage{authblk}
\usepackage{amsmath}
\usepackage{booktabs}
\usepackage{enumitem}

\title{FastSAG: Towards Fast Non-Autoregressive Singing Accompaniment Generation}

\makeatletter
\renewcommand\AB@affilsepx{ \protect\Affilfont} % 这里使用了一个空格来作为分隔符
\renewcommand\AB@affilnote[1]{\mbox{\textsuperscript{#1}}}
\makeatother
\author[1]{Jianyi Chen}
\author[1\dag]{Wei Xue}
\author[2]{Xu Tan}
\author[1]{Zhen Ye}
\author[1]{Qifeng Liu}
\author[1\dag]{Yike Guo}
\affil[1]{
  \protect  \mbox{The Hong Kong University of Science and Technology}
}
\affil[2]{
 \protect \mbox{    Microsoft}
}

\begin{document}

\maketitle

\begin{abstract}
    %Singing accompaniment generation is a task to generate instrumental music to accompany the input vocals, which is crucial to establish a human AI system in symbiotic art creation. The existing method, SingSong\cite{donahue2023SingSong}, employs multi-stage autoregressive model to accomplish singing-to-song generation. However the generation process is slow due to it generates semantic and acoustic tokens one by one in autoregressive manner. We propose FastSAG, the first one to use score-based diffusion model to achieve singing accompaniment generation, which has higher generation speed while maintaining comparable or better quality. To be specific, we employ source separation algorithm on a large audio music dataset to produce a large corpus of vocal and accompaniment pairs. Then the score-based generative model is used to generate accompaniment Mel spectrogram conditioning on vocal audio features. To acquire a better feature of vocal audio, we designed a Semantic Projection Block and a Prior Projection Block to obtain a Mel spectrogram-like prior which will serve as the conditioning feature to the diffusion model. Both objective and subjective evaluation show that our method can achieve better results and higher speed. Audio samples and code are available at \href{www.baidu.com}{www.baidu.com}.

Singing Accompaniment Generation (SAG), which generates instrumental music to accompany input vocals, is crucial to developing human-AI symbiotic art creation systems. The state-of-the-art method, SingSong, utilizes a multi-stage autoregressive (AR) model for SAG, however, this method is extremely slow as it generates semantic and acoustic tokens recursively, and this makes it impossible for real-time applications. In this paper, we aim to develop a Fast SAG method that can create high-quality and coherent accompaniments. A non-AR diffusion-based framework is developed, which by carefully designing the conditions inferred from the vocal signals, generates the Mel spectrogram of the target accompaniment directly. With diffusion and Mel spectrogram modeling, the proposed method significantly simplifies the AR token-based SingSong framework, and largely accelerates the generation. We also design semantic projection, prior projection blocks as well as a set of loss functions, to ensure the generated accompaniment has semantic and rhythm coherence with the vocal signal. By intensive experimental studies, we demonstrate that the proposed method can generate better samples than SingSong, and accelerate the generation by at least 30 times. Audio samples and code are available at \url{https://fastsag.github.io/}.

% \href{https://anonymous.4open.science/r/FastSAG-1623}{this link}.
    
\end{abstract}

\section{Introduction}
Singing Accompaniment Generation (SAG) aims to create instrumental audio tracks that harmonize with vocal performances. This technique empowers individuals to compose complete songs by merely recording their singing. Since the human voice is often regarded as the most intuitive musical instrument, SAG allows people to express their musicality without the need for additional instrumental skills.

% Existing conditional audio music generation tasks mainly focus on text-to-music generation, such as \cite{agostinelli2023musiclm} \cite{schneider2023mo} \cite{huang2023noise2music} \cite{copet2023simple}. These methods may generate instrumental music with singing that are related to the input text prompt. However, they can not guarantee the harmony and consistency between instrumental music and singing. 

Early approaches for SAG are based on retrieval, for instance, the Microsoft Songsmith \cite{simon2008mysong}. The Songsmith extracts pitches of input vocals, then predicts the symbolic chord label sequences that complement the melody, and finally retrieves suitable symbolic instrumental accompaniments from datasets given chord label sequences. An intrinsic limitation of retrieval-based methods is that they actually could not generate new music pieces creatively, therefore, the resulting pieces are not optimal for the vocal inputs. 

Learning-driven approaches are also developed, which generally perform Audio2Audio generation. As a related work, in \cite{wu2022jukedrummer}, the Jukedrummer is proposed to generate a drum audio track based on drumless audio tracks, however, this method cannot perform SAG directly. In \cite{donahue2023SingSong}, the learning-driven SAG is for the first time developed, which, similar to MusicLM \cite{agostinelli2023musiclm} and AudioLM \cite{borsos2023audiolm}, is basically based on autoregressive (AR) language models (LMs) to learn the token associations between the vocals and accompaniments. Multiple LMs are involved in the AR generation, which includes a) semantic LM b) coarse acoustic LM c) fine acoustic LM. The resulting tokens are finally decoded to the audios through the Soundstream \cite{zeghidour2021soundstream} decoder. The main drawback of SingSong is the extremely slow generation speed. Since many AR-based LMs are adopted, the whole generation pipeline becomes complicated and in practice, one second of accompaniment needs dozens of seconds for generation on the Nvidia A100 GPU.

In this paper, we propose FastSAG, a diffusion-based method for fast, coherent, and high-quality SAG. Rather than using AR-based LMs, we design a non-AR diffusion model that directly creates the Mel spectrogram of the accompaniment given specially designed input conditions. In this way, the generation pipeline is substantially simplified and the generation is largely accelerated. To ensure the semantic and rhythm coherence between vocals and accompaniments, when generating the conditions for diffusion, we propose a semantic projection block for semantic alignment, and a prior projection block to enhance the frame-level alignment and control. A set of loss functions are also designed to further improve the semantic and rhythm alignment. Experimental results show that the proposed FastSAG could produce better accompaniments than the SingSong, while accelerating the generation speed by more than 30 times, making the generation to the level of real-time factor smaller than 1. The key contributions are briefly summarized as:
\begin{itemize}[leftmargin=*]
    \item We design a diffusion-based non-AR framework for SAG, which largely simplifies the SAG pipeline as compared with SingSong;
    \item We propose semantic projection block, prior projection block, and a set of loss functions to ensure the rhythmic coherence between vocals and generated accompaniments;
    \item The experimental results demonstrate that the proposed FastSAG significantly accelerates the generation and produces better samples as compared with the baseline.
\end{itemize}

%We re-implement SingSong \cite{donahue2023SingSong} on our own dataset as baseline since the code and pretrained model of SingSong are not released. But there are two differences: using Encodec \cite{defossez2022high} instead of Soundstream \cite{zeghidour2021soundstream} and using MERT \cite{li2023mert} instead of W2v-bert \cite{chung2021w2v}. The reason is that we could not find the open resources of these models. The re-implemented baseline achieved the similar FAD-(vggish \cite{hershey2017cnn}) score on Musdb18 evaluation dataset \cite{rafii2017musdb18} as the original paper reports. 

\begin{figure*}[htb]

\begin{minipage}[b]{1.0\linewidth}
  \centering
  \centerline{\includegraphics[width=16cm]{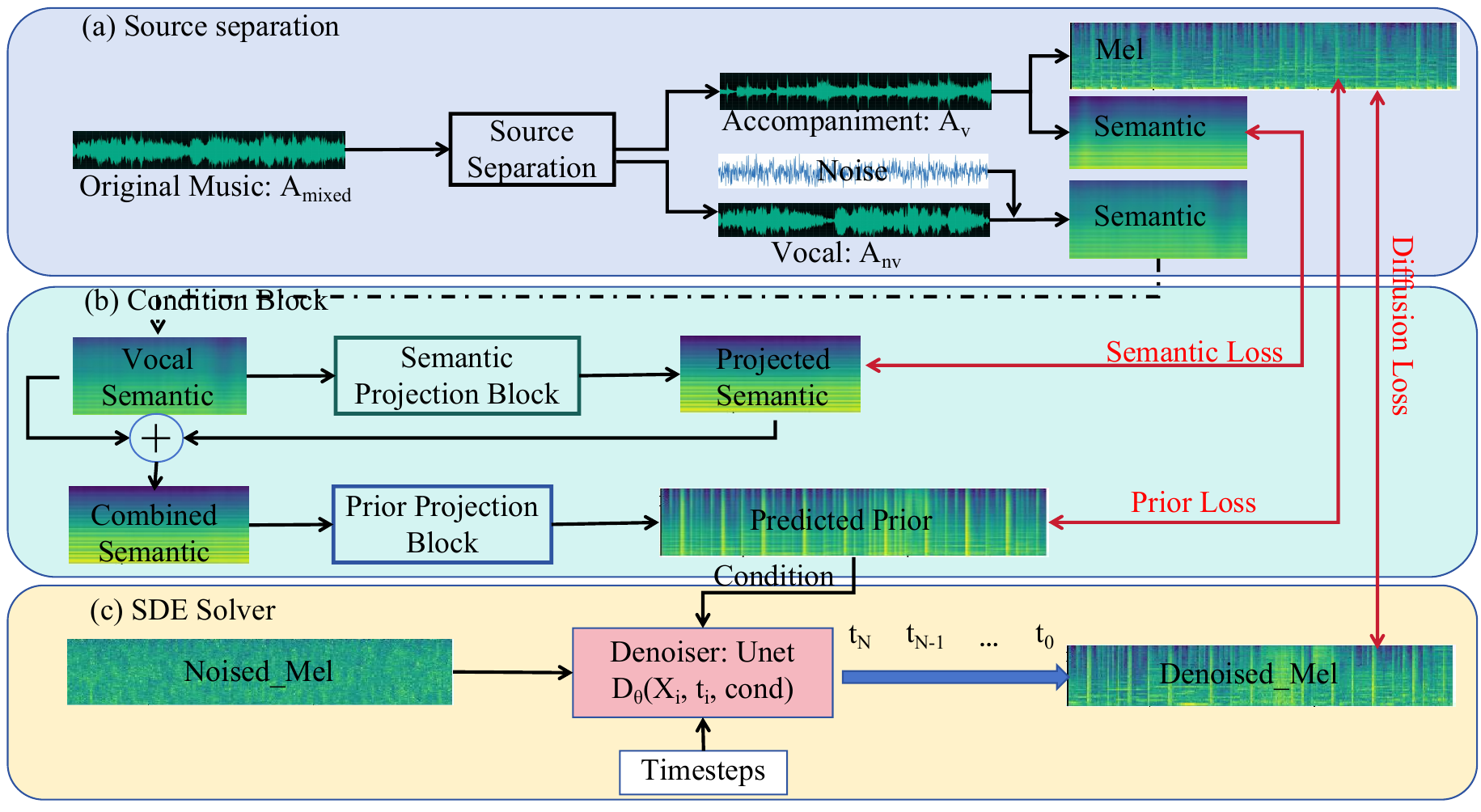}}
%  \vspace{2.0cm}
  %\centerline{(a) Result 1}\medskip
\end{minipage}
\caption{\normalsize Overview of FastSAG. (a) indicates how training data is constructed using a source separation algorithm to acquire vocal-accompaniment pairs. (b) illustrates how to compute conditions based on vocal input. It mainly contains two blocks: semantic projection block and prior projection block. The semantic block is for high-level semantic control and the prior block is for frame-level control. (c) is the stochastic differential equation (SDE) solver, which will take prior computed in (b) as a condition. In the inference process, the generated Mel spectrogram will be converted to an audio waveform through BigvGAN.}
\label{fig1:res}
\end{figure*}

\section{Related Works}
Besides the SAG which we have reviewed in the introduction, here, we further discuss the accompaniment generation given instrumental inputs, and the audio generation methods, which would facilitate the discussions in subsequent sections.
%\subsection{Singing Voice Synthesis}
%Singing voice and instrumental accompaniment consist of two important components of audio music. Singing voice synthesis (SVS) is to generate singing voice with accurate pitch and rhythm taken aligned pitch-lyrics and timbre as inputs, and it has garnered significant attention in recent years due to the increasing applications in interactive AI and entertainment. XiaoiceSing \cite{lu2020xiaoicesing} was among the first of the deep learning driven SVS systems that saw commercial deployment. Further they proposed XiaoiceSing2 \cite{wang2022xiaoicesing} which improves the quality of the singing voice over XiaoiceSing. HiFiSinger \cite{chen2020hifisinger} used three GAN discriminators over three overlapping sub-bands to improve the singing quality. And DiffSinger \cite{liu2022diffsinger} employed denoising diffusion probabilistic model \cite{ho2020denoising} on svs achieving high-fidelity voice. Singing accompaniment generation could be seen as the backend of SVS since it generate the instrumental music to accompany the singing voice to achieve the full music.

\subsection{Accompaniment Generation Given Instrumental Inputs}
For symbolic music, the melody track and the remained accompaniment tracks could be separated easily, and there are some works for symbolic accompaniment generation that take the melody track as input. PopMAG \cite{ren2020popmag} is proposed to generate the accompaniment track which consists of drum, piano, string, guitar, and bass track based on MuMIDI representation, by using Transformer-XL \cite{dai2019transformer} as the backbone. MuseFlow \cite{ding2023museflow} uses the flow model to generate accompaniment based on the revised piano-roll representation. In \cite{wang2022songdriver}, the SongDriver is proposed for real-time accompaniment generation, which consists of two phases: arrangement phase and prediction phase. The above methods generally aim to generate symbolic music. In \cite{mariani2023multi}, by still relying on the symbolic MIDI dataset, the multi-track accompaniment generation is performed in the audio domain based on the rendered multi-track audio waveforms.

\subsection{Audio Generation}
SAG is a type of audio generation task. Significant progress has been made in generating general audio, music, and speech with the advancement of generative models. Now we discuss the methods in terms of representations used: a) raw audio waveform, b) hand-crafted representation, and c) neural representation. 

The raw audio waveform captures the fine-grained details of audio and serves as the direct representation of audio data. Wavenet \cite{oord2016wavenet} employs dilated convolutions to capture long-range dependencies in audio signals, enabling the generation of high-quality and realistic audio. WaveGlow \cite{prenger2019waveglow} achieves high-quality audio synthesis using a flow-based approach, which allows for efficient sampling and parallel processing. Working directly with raw waveforms can be challenging due to their high dimensionality and complex temporal dependencies. 

The most common hand-crafted representation for audio generation is the Mel spectrogram. The audio is generally produced by first relying on an acoustic model to produce the Mel spectrogram given input controls (e.g., text for speech synthesis, and prompts for general audio and music generation), and then a vocoder to convert the Mel spectrogram to the audio domain. For speech synthesis, typical acoustic models include FastSpeech \cite{ren2019fastspeech}, GradTTS \cite{popov2021grad}, and CoMoSpeech \cite{ye2023comospeech}, and for general audio and music generation, acoustic models such as Riffusion\footnote{\url{https://github.com/riffusion/riffusion}}, Mousai \cite{schneider2023mo}, Noise2Music \cite{huang2023noise2music}, Make-An-Audio \cite{huang2023make}, and AudioLDM \cite{liu2023audioldm} are developed. HiFi-GAN \cite{kong2020hifi}, BigvGAN \cite{lee2022bigvgan}, VOCOS \cite{siuzdak2023vocos} are popular vocoders to recover audio waveform. 

Another type of representation is the neural tokens (such as SoundStream \cite{zeghidour2021soundstream}, Encodec \cite{defossez2022high}, DAC \cite{kumar2023high}), which learn a discrete representation from the audio waveform. Further, AR model and LMs could be used to model the evolutions of these discrete tokens in chain rule, which leads to a series of token-based audio generation methods, including unconditional audio generation (audioLM \cite{borsos2023audiolm}), text-to-music generation (MusicLM \cite{agostinelli2023musiclm}, MusicGen \cite{copet2023simple}), text-to-audio (AudioGen \cite{kreuk2022audiogen}), accompaniment generation (SingSong \cite{donahue2023SingSong}), text-to-speech TTS (\cite{wang2023neural}). In addition, the continuous representation from residual VQ (RVQ) codebooks could also be modeled using the diffusion model, for instance, in NaturalSpeech2 \cite{shen2023naturalspeech}.

\section{EDM Formulation}

In this section, we introduce the EDM diffusion model \cite{karras2022elucidating}, which will be used as a conditional probability model illustrated in Figure~\ref{fig1:res} (c). 

Supposing the data distribution is $p_{data}(x)$ and considering the family of mollified distributions $p(x; \delta)$ which is obtained by adding Gaussian noise $N(0, \delta I)$ to the data, the idea of EDM diffusion model is to randomly sample a noisy sample $x_0 \sim N(0, \delta_{max} I)$, and further sequentially denoise it into sample $x_i \sim N(0, \delta_{i} I)$ with noise levels $\delta_{0} = \delta_{max} > \delta_{1} > ... > \delta_{N} = 0$. The number of sampling steps is denoted as $N$, and the final outcome of this process, $x_N$, exhibits a distribution that aligns with the original data. This diffusion process belongs to the variance exploding family \cite{song2020score}.

The stochastic differential equation (SDE) of diffusion is expressed as \cite{song2020score}: 
\begin{align}
    d\mathbf{x} = f(\mathbf{x}, t)dt + g(t)d\mathbf{w},
\end{align}
where $f(\mathbf{x}, t)$ and $g(t)$ mean drift and diffusion coefficients respectively, and $w$ is the standard Wiener process. 
% The probability flow ODE increases the noise level of samples during the forward process and decreases it during the inverse process.

The above process corresponds to a probability flow ordinary differential equation (ODE), and in EDM \cite{karras2022elucidating}, a schedule of $\delta(t)$ is chosen to specify the desired noise level at time $t$. Then the ODE is expressed as
\begin{align}
    d\mathbf{x} = -\dot{\delta}(t)\delta(t)\nabla_{\mathbf{x}}logp(\mathbf{x};\delta(t))dt,
    \label{eq2}
\end{align}%
which defines the process evolving a sample $\mathbf{x}_i \sim p(\mathbf{x}_i; \delta(t_i))$ from time $t_i$ to $t_j$ yields a sample $\mathbf{x}_j \sim p(\mathbf{x}_j; \delta(t_j))$, and $t_i$ to $t_j$ can be either forward or reverse in time. The $\dot{\delta}()$ denotes a time derivative, and $\nabla_{\mathbf{x}}logp(\mathbf{x};\delta(t))$ is the score function \cite{song2020score}. As long as the score function is known, the probability flow ODE in \eqref{eq2} can be used for sampling.

One natural way for $\delta(t)$ scheduling is $\delta(t) \propto \sqrt{t}$, which corresponds to the constant-speed heat diffusion. However, \cite{karras2022elucidating} shows that it is not convenient practically. EDM adopts another schedule, which uses
\begin{align}
    In(\delta) \sim N(P_{mean}, P_{std}^2)
\end{align}%
for training, and 
\begin{align}
    \delta_{i<N}=({\delta_{max}}^{1\over\rho}+{i\over{N-1}}({\delta_{min}}^{1\over\rho}-{\delta_{max}}^{1\over\rho}))^\rho
\end{align}%
for sampling. In our setting of EDM, $P_{mean}=-1.2$, $P_{std}=1.2$, $\delta_{min}=0.002$, $\delta_{max}=80$, and $\rho=7$ which controls how much the steps near $\delta_{min}$ are shortened at the expense of longer steps near $\delta_{max}$.

Supposing $D(\mathbf{x}; \delta)$ is the denoising function that minimizes the expected L2 denoising error for samples drawn from $p_{data}$ separately for every $\delta$, i.e., 
\begin{align}
    E_{\mathbf{y} \sim p_{data}}E_{\mathbf{n} \sim N(0, \delta(t)^2 I)}||D(\mathbf{y+n}; \delta(t)) - \mathbf{y}||_2^2,
\end{align}%
the score function can be written as,
\begin{align}
    \nabla_{\mathbf{x}}logp(\mathbf{x};\delta(t))={D(\mathbf{x}; \delta(t)) - \mathbf{x} \over \delta(t)^2},
\end{align}%,
where $\mathbf{y}$ if the training sample and $\mathbf{n}$ is the noise. 

In the diffusion model, the denoiser $D(\mathbf{x}; \delta(t))$ can be implemented as a neural network $D_{\theta}(\mathbf{x}_t)$ and $D_{\theta}(\mathbf{x}_t, cond)$ for unconditional and conditional diffusion respectively, where $cond$ is the condition. Similar to the EDM setting, 
\begin{align}
    D_{\theta}(\mathbf{x}_t, cond) = c_{skip}(t)\mathbf{x}_t + c_{out}F_{\theta}(\mathbf{x}_t, t, cond),
\end{align}%
where $F_{\theta}$ can be any well-designed neural network, for example, Wavenet \cite{oord2016wavenet} is used in \cite{liu2022diffsinger} and U-net \cite{ronneberger2015u} is used in \cite{popov2021grad}. $c_{skip}$ and $c_{out}$ are used to control the skip connect and the magnitudes of $F_{\theta}$, respectively. The $c_{skip}$ and $c_{out}$ can be expressed as
\begin{align}
    c_{skip}(t)={\delta_{data}^2 \over {(t-\epsilon)^2 + \delta_{data}^2}}, c_{out}(t)={\delta_{data}(t-\epsilon) \over {\sqrt{\delta_{data}^2 + t^2}}},
\end{align}%
where $\delta_{data}=0.5$ and $\epsilon=0.002$ denoting the smallest time interval during sampling.
\section{Proposed FastSAG}
In this section, we introduce our proposed method FastSAG for singing accompaniment generation. As shown in Figure.~\ref{fig1:res}, it contains three main parts: source separation for data processing, condition block for computing condition of the diffusion model, and EDM-based SDE solver for generating the Mel spectrogram of accompaniment. 

\subsection{Overview}
Most public songs on the internet are audio mixes of vocals and accompaniments, and we denote the mixture signal as $A_{mixed}$, the vocal as $ A_{v}$, and the accompaniment as $A_{nv}$ respectively. To obtain the paired vocal and accompaniment data, source separation, such as Demucs \cite{defossez2019demucs}, is applied to the mixture signals, and resulting pseudo vocal-accompaniment pairs $(A_{v}, A_{nv})$ are acquired. Similar to SingSong \cite{donahue2023SingSong}, slight white Gaussian noise $N_{oise}$ is added to the vocal input to mitigate source separation artifacts. The core of SAG is building the conditional probabilistic model $P(A_{nv}|A_{v}+N_{oise})$. 

Different from SingSong which uses the LM for discrete audio tokens and is time-consuming, we design a diffusion-based non-AR framework here. In the continuous Mel spectrogram space, the EDM introduced in Section 3 takes the conditions containing the semantic and rhythmic information to generate the Mel spectrogram of the accompaniment, denoted as ${Mel^{\prime}_{nv}}$. The resulting Mel spectrogram is transformed to the audio domain by using Bigvgan \cite{lee2022bigvgan} as a vocoder. 
\subsection{Condition Block}

\begin{figure}[htb]
\begin{minipage}[b]{1.0\linewidth}
  \centering
  \centerline{\includegraphics[width=9cm]{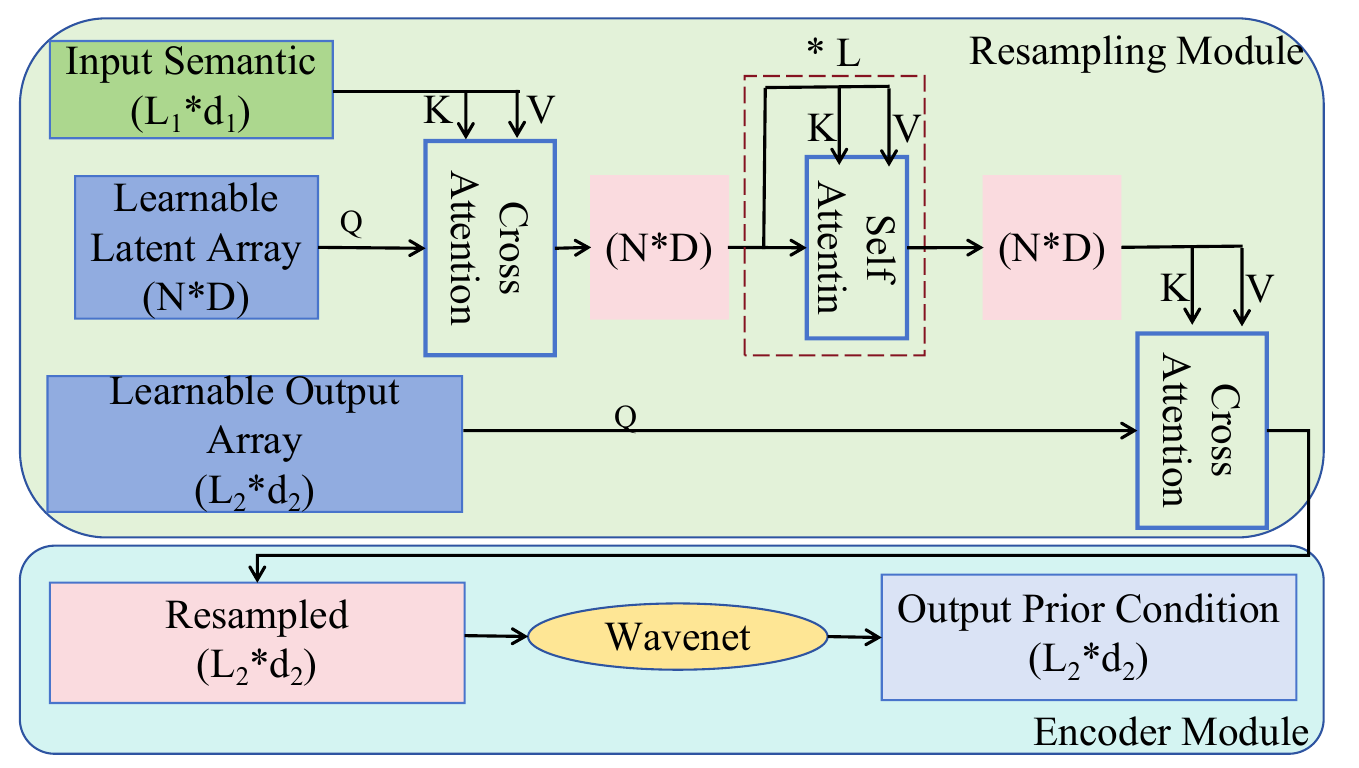}}
%  \vspace{2.0cm}
  %\centerline{(a) Result 1}\medskip
\end{minipage}
\caption{\normalsize The Prior Projection Block. It contains one resampling module and one encoder module. The resampling module is for reshaping the feature shape, mapping from semantic feature shape to Mel. And encoder module is further for prior generation.}
\label{fig2:res}
\end{figure}

%In order to generate coherent and harmonious instrumental music to accompany singing audio, how to compute the condition becomes significant. The coherence and harmony are important both in overall-level and frame-level. 
%We design two cascading blocks: Semantic Projection Block used for mapping high-level semantic feature and Prior Projection Block used for generating frame-level condition. The insight for cascading design is that modeling the frame-level acoustic relationship directly is difficult. So we model the high-level relationship is easier. 

The condition block takes the vocal signal as input, and incorporates two cascading blocks: a) the Semantic Projection Block, which is responsible for mapping high-level semantic features, and b) the Prior Projection Block, which generates frame-level aligned conditions. The rationale behind this cascading design is that directly modeling the frame-level acoustic relationship is challenging, while modeling the high-level relationship is comparatively easier. Hence, by first capturing the high-level semantic features and then generating frame-level conditions based on them, we can effectively address the complexity of modeling the frame-level acoustic relationship.

\subsubsection{Semantic Projection Block}
Denoting the input vocal audio signal as $A_{v} \in R^{T\times 1}$ with $T$ as the number of frames in waveform format, the high-level semantic feature $S_v \in R^{L_1\times d_1}$ is extracted using MERT \cite{li2023mert}, where $L_1$ is the frame number of MERT feature and $d_1$ is the dimension. The Semantic Projection Block consists of a neural network (for example, Wavenet \cite{oord2016wavenet}) to obtain the predicted semantic feature of the accompaniment $S_{nv}^{\prime} =\texttt{Wavenet}(S_v)\in R^{L_1\times d_1}$. 

\subsubsection{Prior Projection Block}
A high-level semantic feature is utilized for generating frame-level roughly aligned prior through the Prior Projection Block further. Inspired by Grad-TTS \cite{popov2021grad}, they used the frame-aligned prior as the condition to diffusion model by using an aligner to transform phoneme-level feature to frame-level prior. Illustrated as Figure \ref{fig2:res}, our Prior Projection Block consists of a resampling module and an encoder module.

Considering that the shape (time resolution and feature dimension) of the semantic feature may differ from that of the desired frame-level prior, we design two kinds of resampling modules. The first one is using bi-linear interpolation directly. The second one is that we utilize Perceiver-IO \cite{jaegle2021perceiver} as the resampling module to obtain the resampled semantic feature ${R}_{nv}^{\prime}=\texttt{PerceiverIO}(S_{nv}^{\prime}+S_v) \in R^{L_2 \times d_2}$, where we mixed the vocal semantic feature $S_v$ and predicted accompaniment semantic $S_{nv}^{\prime}$ by just adding to acquire better performance. $L_2$ and $d_2$ denote the length and number bin of Mel spectrogram, respectively.

The computation of $\texttt{PerceiverIO}$ is described as follows. The input of Perceiver-IO $x =S_{nv}^{\prime}+S_v\in R^{L_1 \times d_1}$ is then encoded into a latent space by cross attention, yielding $z_1=\texttt{crossAttn}(x,l) \in R^{N \times D}$, where  $l \in R^{N \times D}$ is a learnable variable. Multiple self-attention layers further convert $z_1$ to another hidden embedding $z_2=\texttt{multiSelfAttn}(z_1) \in R^{N \times D}$, and then $z_2$ is decoded to output array $z_3=\texttt{crossAttn}(o,z_2) \in R^{L_2 \times d_2}$ with cross attention, where $o \in R^{L_2 \times d_2}$ is a learnable output query. The output array $z_3$ is the desired resampled semantic feature ${R}_{nv}$ introduced. In our experiments, we set $N=32, D=256$.

Further, the resampled semantic feature ${R}_{nv}$ is processed to obtain a Mel spectrogram-like prior $P_{rior}=\texttt{Wavenet}({R}_{nv}^{\prime}) \in R^{L_2 \times d_2}$ through encoder module, which serves as the condition for diffusion model latter. We use Wavenet \cite{oord2016wavenet} as the encoder module again.

\subsection{Conditional Denoiser}
Similar to Grad-TTS \cite{popov2021grad}, we use Unet2d \cite{ronneberger2015u} as denoiser, but with a slight modification. As introduced in Section 3, different level of noise is sampled from $\ln(\delta) \sim N(P_{mean}, P_{std}^2)$, we have no explicit timesteps $t$ controlling noise level. Therefore, we replace $t$ with $\ln(\delta)$ and take it as one of the inputs to denoiser $D_{noised}Mel_{nv}=D_{\theta}(N_{oised}Mel_{nv}, \ln(\delta), P_{rior})$, where $N_{oised}Mel_{nv} \in R^{L_2 \times d_2}$ is the noised Mel spectrogram of accompaniment, ${R}_{nv}^{\prime}$ is the condition computed in last subsection, and $D_{noised}Mel_{nv} \in R^{L_2 \times d_2}$ is the denoised output. 

\subsection{Loss Function}
To make the model generate coherent and harmonious accompaniments, by taking the separated accompaniment $A_{nv}\in R^{T\times1}$ as ground truths, we design three different loss functions, including the semantic loss, prior loss, and diffusion loss. We extract semantic ground truths $S_{nv} \in R^{L_1 \times d_1}$ and Mel spectrogram ground truths $Mel_{nv} \in R^{L_2 \times d_2}$ of accompaniment music through MERT \cite{li2023mert} and Bigvgan \cite{lee2022bigvgan}, respectively, the loss functions are computed as below.

\textbf{Semantic Loss}. The goal of the semantic loss is to establish a high-level semantic relationship between the vocal and accompaniment tracks. We believe it is much easier than constructing the acoustic mapping directly, and it is defined as:
\begin{align}
    L_{semantic} = ||S_{nv}^{\prime}-S_{nv}||_2^2.
\end{align}%
\textbf{Prior Loss}. The purpose of the prior loss is to establish a rough frame-level alignment between the Mel spectrogram of the accompaniment track. It is defined as,
\begin{align}
    L_{prior} = ||P_{rior}-Mel_{nv}||_2^2.
\end{align}%
\textbf{Diffusion Loss}. While the condition block alone can generate a rough accompaniment, the diffusion model is employed to produce a refined accompaniment by conditioning it on the prior. The diffusion loss is defined as,
\begin{align}
    L_{diffusion} = ||D_{noised}Mel_{nv}-Mel_{nv}||_2^2.
\end{align}%

The final loss function is a combination of semantic loss, prior loss, and diffusion loss, as
\begin{align}
    L=L_{semantic} \times \lambda_{s} + L_{prior} \times \lambda_{p} + L_{diffusion} \times \lambda_{d},
\end{align}%
where $\lambda_{s}$, $\lambda_{p}$ and $\lambda_{d}$ are the weights of corresponding loss terms. In our experiments, we set $\lambda_{s}=1.0$, $\lambda_{p}=1.0$ and $\lambda_{d}=1.0$.

\section{Experiments}
% In this section, we introduce data processing, evaluation metrics, implementation details, and experiment results.
\subsection{Dataset}
Here, we discuss the data collection and processing, as well as the composition of the training and evaluation datasets.

\begin{table*}
    \centering
    \resizebox{\textwidth}{!}{
        \begin{tabular}{lrrrrrrr}
            \toprule
            Methods  & $\text{FAD}_{\text{VGGish}}$($\downarrow$) & $\text{FAD}_{\text{MERT}}$($\downarrow$) & $\text{FAD}_{\text{MERT}_4}$($\downarrow$) & $\text{FAD}_{\text{MERT}_7}$($\downarrow$) & $\text{FAD}_{\text{MERT}_{11}}$($\downarrow$) & $\text{FAD}_{\text{CLAP-MUSIC}}$($\downarrow$) & RTF($\downarrow$) \\
            \midrule
            $\text{SingSong}_{s+c}$     & 0.8632        & 3.1589     & 1.7528 & 1.5793 & 1.6917 & 0.0878 & 10.4936  \\
            $\text{SingSong}_{s+c+f}$     & 1.5578          & 3.7985    & 2.3694 & 2.0688 & 2.1319 & 0.1363  & 47.7660  \\
            \bottomrule
            FastSAG(w/o norm)      & 3.6784  & 1.7695  & 1.7142 & 1.5849 & 1.2951 & 0.1115 & 0.3239  \\
            FastSAG(seman+mel)      & 1.5947     & 1.9795    & 1.5769 & 1.8818 & 1.5152 & 0.1054  & 0.3240  \\
            FastSAG(mel)      & 1.4424     & 2.0861    & 1.6532 & 1.8578 & 1.5370 & 0.1151  & \textbf{0.3214}  \\
            FastSAG(interpolate)      & \textbf{0.7595}     & 1.5059    & 1.0806 & 1.3566 & 1.0629 & \textbf{0.0648}  & 0.3231  \\
            FastSAG          & 0.8917      & \textbf{1.3043}  & \textbf{0.9675} & \textbf{1.1364} & \textbf{0.8227} & 0.0701  & 0.3247    \\
        \end{tabular}
    }
    \caption{\normalsize FAD on zero-shot MUSDB18 test dataset. $\text{SingSong}_{s+c}$ means only using semantic model and coarse acoustic model. $\text{SingSong}_{s+c+f}$ means using the semantic model, coarse acoustic model, and fine acoustic model. FastSAG(w/o norm) means using the original logarithmic Mel spectrogram instead of the normalized one. FastSAG(seman+mel) means using the semantic feature and Mel spectrogram of vocal audio as a condition. FastSAG(mel) means only using the Mel spectrogram of vocal audio as a condition. FastSAG(interpolate) means using the interpolate operator as a re-sampling module in the prior projection block instead of Perceiver-IO. And FastSAG is the method introduced in the previous section which only uses the semantic feature of vocal audio as a condition.}
    \label{tab:tbl1}
\end{table*}

\begin{table*}
    \centering
    \resizebox{14cm}{!}{
        \begin{tabular}{lrrrrrr}
            \toprule
            Methods  & $\text{FAD}_{\text{VGGish}}$($\downarrow$) & $\text{FAD}_{\text{MERT}}$($\downarrow$) & $\text{FAD}_{\text{MERT}_4}$($\downarrow$) & $\text{FAD}_{\text{MERT}_7}$($\downarrow$) & $\text{FAD}_{\text{MERT}_{11}}$($\downarrow$) & $\text{FAD}_{\text{CLAP-MUSIC}}$($\downarrow$) \\
            \midrule
            $\text{SingSong}_{s+c}$     & \textbf{0.5093}          & 1.2798     & 0.7691 & 0.8144 & 0.7832 & 0.0406 \\
            \bottomrule
            
            FastSAG(interpolate)     & 1.3266    & \textbf{0.6450}  & 0.5101 & 0.6267 & 0.4676 & 0.0361    \\
            FastSAG        & 1.2180     & 0.6639  & \textbf{0.4295} & \textbf{0.4966} & \textbf{0.4164} & \textbf{0.0330}    \\
        \end{tabular}
    }
    \caption{\normalsize FAD on in-domain test dataset. }
    \label{tab:tbl2}
\end{table*}

\textbf{Data Collection and Processing}. We collected over 300k songs from public sources on the internet, most of which are Chinese and English songs. By using Demucs \cite{defossez2019demucs}, we obtain mono vocal-accompaniment pairs with a sampling rate of 44.1~kHz. Then, filtering is applied to keep sample pairs with clear vocals and accompaniments, which is done by first obtaining 10-second paired clips and then keeping pairs where both the vocal and accompaniment have a peak RMS amplitude over -25dB. As a result, we obtain a collection of over 1.2 million pairs of 10-second clips, totaling more than 3000 hours.

\textbf{Training Dataset}. We divide the 1.2 million pairs of 10-second clips into training and (in-domain) evaluation datasets, with 2,000 samples for in-domain evaluation. To ensure that the evaluation samples are not seen during training, samples of each song are exclusively allocated to either the training dataset or the evaluation dataset.

\textbf{Evaluation Dataset}. We construct two kinds of evaluation datasets: the in-domain evaluation dataset, which has been introduced, and the zero-shot evaluation dataset. Although the in-domain evaluation dataset is not used for training, they follow a similar distribution. An out-of-domain zero-shot evaluation dataset is additionally built from the MUSDB18 dataset \cite{rafii2017musdb18}, which is also the evaluation dataset of SingSong \cite{donahue2023SingSong}. Following the same procedures for training data processing on the MUSDB18 test dataset, we obtain 348 paired clips. The MUSDB18 training dataset is not used for training so the MUSDB18 test dataset is zero-shot. 

\subsection{Baselines and Implementation Details}
%We design two baseline methods: SingSong and RandSong. Since we can not find official resource codes of these methods, here we introduce the implementation details of SingSong and RandSong, as well as our FastSAG. SingSong and FastSAG are trained on the same dataset. 

We use two baseline methods, SingSong and RandSong, for comparison. As we could not find source codes for baselines, we provided implementation details for SingSong, RandSong, and the proposed FastSAG. Both SingSong and FastSAG are trained on the same dataset to ensure a fair comparison.

\textbf{SingSong}. We implemented SingSong based on the codebase of open-musiclm \footnote{\url{https://github.com/zhvng/open-musiclm.git}}. Same with open-musiclm, we utilize  Encodec \cite{defossez2022high} as a replacement for SoundStream \cite{zeghidour2021soundstream}, and MERT \cite{li2023mert} as a replacement for w2v-BERT \cite{chung2021w2v}. The main reason for replacement is that we can not find open source code and pre-trained model of SoundStream and w2v-BERT when we re-implement SingSong. We trained the models of the semantic stage, coarse acoustic stage, and fine acoustic stage separately for 240k, 130k, and 240k steps respectively on a single NVIDIA A100 GPU with 80GB memory. The batch size and grad accumulation steps are (24, 1) for the semantic stage model, (12, 4) for the coarse acoustic model, and (4, 4) for the fine acoustic model. The Encodec \cite{defossez2022high} employs an RVQ scheme that generates 8-dimensional acoustic codes at a rate of 75~Hz. The first 3 dimensions correspond to the coarse acoustic codes, while the remaining 5 dimensions represent the fine acoustic codes.

\textbf{RandSong}. The RandSong is a weak baseline to examine the importance and sensitiveness of harmony and coherence between vocal voice and instrumental accompaniment audio. Firstly, we construct a big candidate set of instrumental accompaniment audio (20,000 pieces) from human-composed music separated by Demucs \cite{defossez2019demucs}. Then for a given vocal voice, we randomly choose one accompaniment from the candidate set. The samples from RandSong are only used for subjective evaluation. 

\textbf{FastSAG}. We trained our FastSAG on a single NVIDIA A100 GPU with 80GB memory for 0.5M steps using Adam optimizer with a constant learning rate of 0.0001 and batch size of 28. For the vocoder, we utilize Bigvgan \cite{lee2022bigvgan}, which operates at a sampling rate of  24kHz. The Mel spectrogram used by Bigvgan consists of 100 bins and is computed at a rate of 93.75Hz. We normalize the logarithmic Mel spectrogram to -1 to 1, instead of using the original one, because the original logarithmic Mel spectrogram ranges from -12 to 2, which is more difficult to model. The normalization and de-normalization are described as:
\begin{align}
    X_{nor} =  {(X - X_{min}) \over {X_{max} - X_{min}}} \times 2 - 1
\end{align}%
\begin{align}
    X_{den} =  {(X_{nor} + 1) \over 2} \times ({X_{max} - X_{min}}) + X_{min}
\end{align}%
where $X$, $X_{nor}$ and $X_{den}$ denote the original spectrogram, normalized spectrogram and de-normalized spectrogram, respectively, $X_{max}=2$, $X_{min}=-12$ are constants. During the inference process, we employ a first-order ODE solver with a total of 50 sampling steps.

\subsection{Evaluation Metrics}
We examine the proposed model and baselines through objective evaluation and subjective evaluations.

\textbf{Objective evaluation}. SingSong employed FAD-{vggish} as an objective evaluation metric, utilizing embeddings from the VGGish audio classifier \cite{hershey2017cnn}. However, relying solely on FAD-{vggish} may not accurately reflect the true quality of the generated music. 

To address this, we utilize the FADTK \cite{gui2023adapting}, an extended FAD toolkit that includes various embedding extractors specifically designed for evaluating generative music. We incorporate three types of embedding extractors: VGGish \cite{hershey2017cnn}, MERT \cite{li2023mert}, and clap-laion-music \cite{wu2023large}. The latter two are trained on large music datasets, resulting in the evaluation metrics $\texttt{FAD}_{\texttt{VGGish}}$, $\texttt{FAD}_{\texttt{MERT}}$, and $\texttt{FAD}_{\texttt{CLAP-MUSIC}}$. For MERT embeddings, we conduct experiments using different layers, specifically the 4th, 7th, and 11th layers, denoted as $\texttt{FAD}_{\texttt{MERT-4}}$, $\texttt{FAD}_{\texttt{MERT-7}}$, and $\texttt{FAD}_{\texttt{MERT-11}}$, respectively. 

When calculating FADs, ground truth mixtures may be degraded using the corresponding vocoder or codec, to eliminate the impact of sampling rate, vocoders, and codecs on the data distribution.

In addition to evaluating the quality of the generated music, we also assess the speed of the music generation process using the real-time factor (\texttt{RTF}), which is calculated as the ratio between the total time taken for audio generation and the duration of the generated audio.

%\textbf{Subjective Evaluation}. We use MOS to measure the quality of generated music by human evaluation. We invite 15 listeners to rate the harmony and coherence of the testing samples on a scale of range 1 to 5. The testing samples involving mixture composed by human, mixture generated by SingSong, mixture with random selected accompaniment (denoting as RandSong), and mixture generated by FastSAG. 

\textbf{Subjective evaluation}. To assess the quality of the generated music, we employ Mean Opinion Score (MOS) through human evaluation. We invited professional 15 listeners to rate the harmony and coherence of the testing samples on a scale ranging from 1 to 5. The testing samples include mixtures composed by humans, mixtures generated by SingSong, mixtures with randomly selected accompaniments (referred to as RandSong), and mixtures generated by FastSAG.

\subsection{Experiment Results}

\begin{table}
    \centering
    \resizebox{5cm}{!}{
        \begin{tabular}{lr}
            \toprule
            Methods  & $\text{MOS}$($\uparrow$) \\
            \midrule
            Human-composed       & 4.15\\
            $\text{SingSong}_{s+c}$     & 2.36\\
            RandSong       & 1.48 \\
            
            \bottomrule
            
            $\text{FastSAG}_\text{interpolate}$         & 2.78 \\
            FastSAG          & \textbf{3.13} \\
        \end{tabular}
    }
    \caption{\normalsize  Human subjective evaluation on harmony and coherence. Testing samples are chosen from the in-domain test dataset and zero-shot MUSDB18 test dataset with a ratio 2:1.}
    \label{tab:tbl3}
\end{table}

\subsubsection{A. Objective Evaluation}
Table~\ref {tab:tbl1} shows the objective evaluation results on the zero-shot MUSDB18 test dataset, including comparison with baseline SingSong and some ablation studies. 

\textbf{Zero-shot Evaluation}. For baseline SingSong, we analyze two-stage (semantic + coarse acoustic) and three-stage (semantic + coarse acoustic + fine acoustic) inference results. We find two-stage results are better than three-stage ones both in FAD and RTF metrics in our re-implementation, and the best $\text{FAD}_\text{VGGish}$ 0.8632 is better than the original paper reported (0.96). Our method FastSAG with bi-linear interpolation as a resampling module achieves better results both in all FAD metrics and over 30 times faster than two-stage SingSong and over 140 times faster than three-stage SingSong. Our method FastSAG with bi-linear interpolation as the resampling module results in improved performance across all FAD metrics. Additionally, it achieves a significant acceleration, being over 30 times faster than the two-stage SingSong method and over 140 times faster than the three-stage SingSong method. 

\textbf{Ablation Study}. We also conduct several ablation studies to check the function of different settings, consisting of three aspects: 
\begin{itemize}[leftmargin=*]
\item Normalization is important. Using a normalized Mel spectrogram achieves better performance than an unnormalized one in all FAD metrics.
\item Semantic as the condition is better. We conduct experiments with different condition types: only semantic feature, only Mel spectrogram feature, and mixed feature of semantic and Mel spectrogram feature. We can see using only the semantic feature as the condition achieves better performance in all FAD metrics, which is consistent with the findings of SingSong.
\item Discussion of resampling module. Our findings indicate that the utilization of bi-linear interpolation leads to improved FAD scores in $\text{FAD}_\text{VGGish}$ and $\text{FAD}_\text{CLAP-MUSIC}$. However, the FAD scores are comparatively lower when using MERT as the feature extractor. To further assess their performance, we will conduct subjective evaluations.
\end{itemize}

\textbf{In-Domain Evaluation}. We additionally assess the performance of the in-domain test dataset. As shown in Table~\ref{tab:tbl2}, SingSong achieves the highest $\text{FAD}\text{VGGish}$ score, but it performs relatively worse in terms of FAD score when using other feature extractors. While our method utilizing interpolation as the resampling module achieves improved $\text{FAD}\text{VGGish}$ and $\text{FAD}_\text{CLAP-MUSIC}$ compared to the approach employing Perveiver-IO as the resampling module, it exhibits poorer performance on the in-domain test dataset.

\subsubsection{B. Subjective Evaluation}
Table~\ref{tab:tbl3} presents the subjective evaluation results. For this evaluation, 100 testing samples were randomly chosen from both the in-domain test dataset and the zero-shot MUSDB18 test dataset, maintaining a 2:1 ratio respectively. We engaged 15 participants in the evaluation process, with each individual assessing 20 samples randomly picked from the 100 selected samples. The results indicate that RandSong received the lowest MOS score, highlighting the importance of harmony and coherence in human music perception. Furthermore, our FastSAG method, whether employing interpolation or the Perceiver-IO as the resampling module, consistently outperforms the baseline SingSong. However, using the Perceiver-IO as the resampling module results in the closest approximation to human-composed music.

\begin{figure}[htb]
\begin{minipage}[b]{1.0\linewidth}
  \centering
  \centerline{\includegraphics[width=9cm]{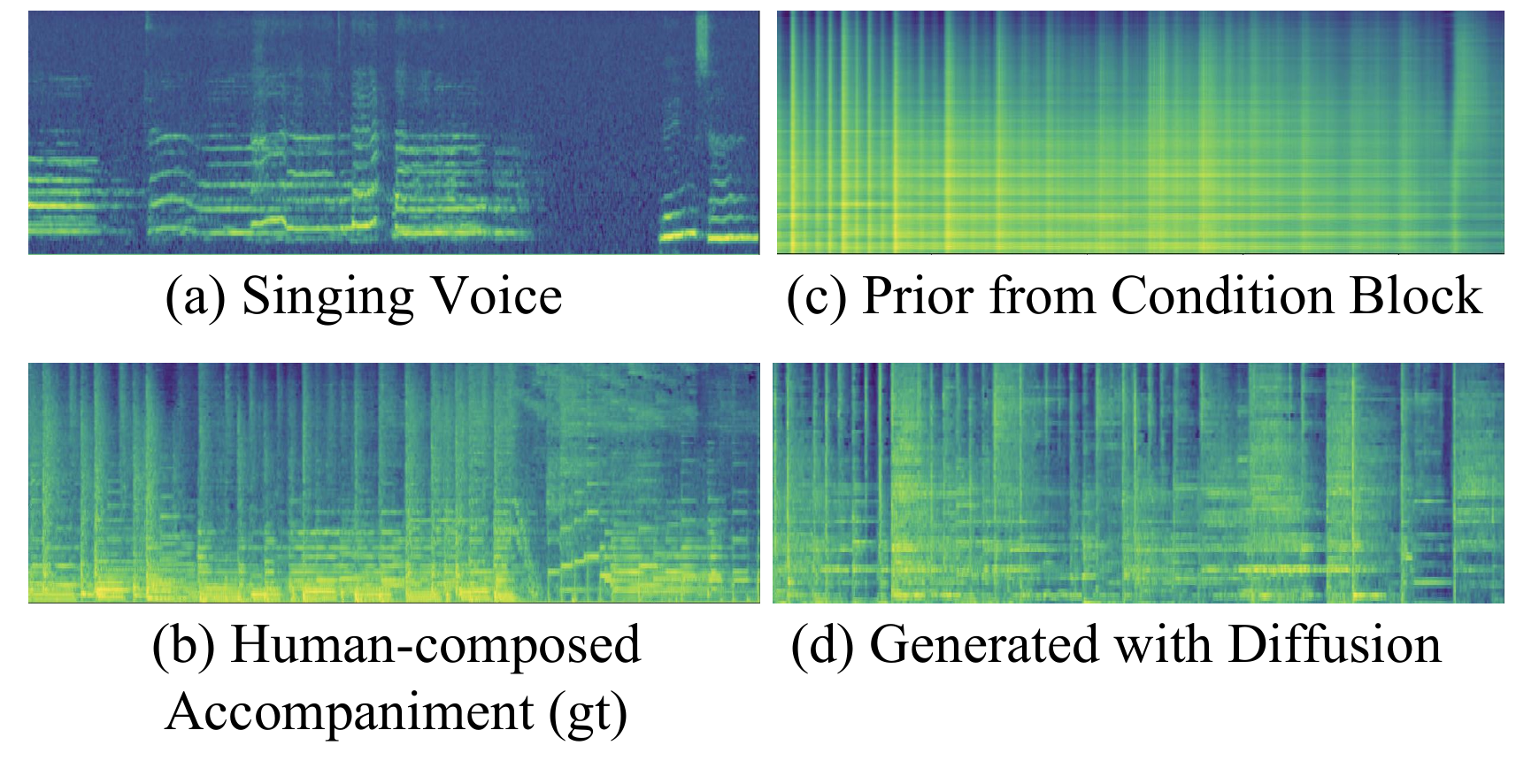}}
%  \vspace{2.0cm}
  %\centerline{(a) Result 1}\medskip
\end{minipage}
\caption{\normalsize Illustration of Mel spectrogram. (a) is Mel of the singing voice. (b) is human-composed accompaniment serving as our growth-truth. (c) is predicted prior. (d) is the generated Mel from the diffusion model.}
\label{fig3:res}
\end{figure}

\subsubsection{C. Discussion on Prior Loss and Diffusion Loss}
Figure \ref{fig3:res} (a) and (b) display the Mel spectrograms of the singing voice and corresponding accompaniment, respectively. There is a significant gap between their Mel spectrograms. (c) represents the predicted prior, which is a first step towards being closely aligned with the accompaniment. As seen in the figure, it is roughly aligned. (d) is generated by the diffusion model, conditioned on (c), and presents a more detailed approximation of the accompaniment. In summary, our framework consists of two stages. The first stage involves a single encoder-decoder, which is not sufficient for generating accompaniment. The second stage refines the output using a diffusion model. The complete framework serves as a conditional probabilistic model for the complex mapping between singing voice and accompaniment.

\section{Conclusion and Future Work}
 In this paper, we introduce FastSAG, a diffusion-based non-autoregressive method for singing accompaniment generation. This approach not only achieves higher generation speeds but also maintains more harmonious and coherent accompaniments, as demonstrated by both objective and subjective evaluations. 

There are, however, areas for improvement. First, the generated audio quality could be better. The low audio quality may be due to several factors: a) the low sampling rate of the training data; b) the degradation of audio quality caused by source separation; and c) the potential for vocoders to degrade audio quality. Second, the generated accompaniment consists of numerous instrumental components, so it may be beneficial to generate each track with more fine-grained control. Lastly, both SingSong and our FastSAG are offline algorithms, meaning that the accompaniment is generated for an entire piece of singing voice. In the future, we could explore designing a framework for online accompaniment generation that adapts as the singing voice progresses.

\section*{Acknowledgments}

The research was supported by Early Career Scheme (ECS-HKUST22201322), Theme-based Research Scheme (T45-205/21-N) from Hong Kong RGC, NSFC (No. 62206234), and Generative AI Research and Development Centre from InnoHK.

\bibliographystyle{named}
\bibliography{ijcai24}

\begin{thebibliography}{}

\bibitem[\protect\citeauthoryear{Agostinelli \bgroup \em et al.\egroup }{2023}]{agostinelli2023musiclm}
Andrea Agostinelli, Timo~I Denk, Zal{\'a}n Borsos, Jesse Engel, Mauro Verzetti, Antoine Caillon, Qingqing Huang, Aren Jansen, Adam Roberts, Marco Tagliasacchi, et~al.
\newblock Musiclm: Generating music from text.
\newblock {\em arXiv preprint arXiv:2301.11325}, 2023.

\bibitem[\protect\citeauthoryear{Borsos \bgroup \em et al.\egroup }{2023}]{borsos2023audiolm}
Zal{\'a}n Borsos, Rapha{\"e}l Marinier, Damien Vincent, Eugene Kharitonov, Olivier Pietquin, Matt Sharifi, Dominik Roblek, Olivier Teboul, David Grangier, Marco Tagliasacchi, et~al.
\newblock Audiolm: a language modeling approach to audio generation.
\newblock {\em IEEE/ACM Trans. Audio, Speech, Lang. Process.}, 2023.

\bibitem[\protect\citeauthoryear{Chung \bgroup \em et al.\egroup }{2021}]{chung2021w2v}
Yu-An Chung, Yu~Zhang, Wei Han, Chung-Cheng Chiu, James Qin, Ruoming Pang, and Yonghui Wu.
\newblock W2v-bert: Combining contrastive learning and masked language modeling for self-supervised speech pre-training.
\newblock In {\em Proc. IEEE Workshop on Automatic Speech Recognition and Understanding ({ASRU})}, 2021.

\bibitem[\protect\citeauthoryear{Copet \bgroup \em et al.\egroup }{2023}]{copet2023simple}
Jade Copet, Felix Kreuk, Itai Gat, Tal Remez, David Kant, Gabriel Synnaeve, Yossi Adi, and Alexandre D{\'e}fossez.
\newblock Simple and controllable music generation.
\newblock {\em arXiv preprint arXiv:2306.05284}, 2023.

\bibitem[\protect\citeauthoryear{Dai \bgroup \em et al.\egroup }{2019}]{dai2019transformer}
Zihang Dai, Zhilin Yang, Yiming Yang, Jaime Carbonell, Quoc~V Le, and Ruslan Salakhutdinov.
\newblock Transformer-{xl}: Attentive language models beyond a fixed-length context.
\newblock In {\em Proc. Assoc. for Computational Linguistics (ACL)}, 2019.

\bibitem[\protect\citeauthoryear{D{\'e}fossez \bgroup \em et al.\egroup }{2019}]{defossez2019demucs}
Alexandre D{\'e}fossez, Nicolas Usunier, L{\'e}on Bottou, and Francis Bach.
\newblock Demucs: Deep extractor for music sources with extra unlabeled data remixed.
\newblock {\em arXiv preprint arXiv:1909.01174}, 2019.

\bibitem[\protect\citeauthoryear{D{\'e}fossez \bgroup \em et al.\egroup }{2022}]{defossez2022high}
Alexandre D{\'e}fossez, Jade Copet, Gabriel Synnaeve, and Yossi Adi.
\newblock High fidelity neural audio compression.
\newblock {\em arXiv preprint arXiv:2210.13438}, 2022.

\bibitem[\protect\citeauthoryear{Ding and Cui}{2023}]{ding2023museflow}
Fanyu Ding and Yidong Cui.
\newblock Museflow: music accompaniment generation based on flow.
\newblock {\em Applied Intelligence}, 2023.

\bibitem[\protect\citeauthoryear{Donahue \bgroup \em et al.\egroup }{2023}]{donahue2023SingSong}
Chris Donahue, Antoine Caillon, Adam Roberts, Ethan Manilow, Philippe Esling, Andrea Agostinelli, Mauro Verzetti, Ian Simon, Olivier Pietquin, Neil Zeghidour, et~al.
\newblock Singsong: Generating musical accompaniments from singing.
\newblock {\em arXiv preprint arXiv:2301.12662}, 2023.

\bibitem[\protect\citeauthoryear{Gui \bgroup \em et al.\egroup }{2023}]{gui2023adapting}
Azalea Gui, Hannes Gamper, Sebastian Braun, and Dimitra Emmanouilidou.
\newblock Adapting frechet audio distance for generative music evaluation.
\newblock {\em arXiv preprint arXiv:2311.01616}, 2023.

\bibitem[\protect\citeauthoryear{Hershey \bgroup \em et al.\egroup }{2017}]{hershey2017cnn}
Shawn Hershey, Sourish Chaudhuri, Daniel~PW Ellis, Jort~F Gemmeke, Aren Jansen, R~Channing Moore, Manoj Plakal, Devin Platt, Rif~A Saurous, Bryan Seybold, et~al.
\newblock Cnn architectures for large-scale audio classification.
\newblock In {\em Proc. IEEE Intl. Conf. Acoustics, Speech, Signal Process. (ICASSP)}, 2017.

\bibitem[\protect\citeauthoryear{Huang \bgroup \em et al.\egroup }{2023a}]{huang2023noise2music}
Qingqing Huang, Daniel~S Park, Tao Wang, Timo~I Denk, Andy Ly, Nanxin Chen, Zhengdong Zhang, Zhishuai Zhang, Jiahui Yu, Christian Frank, et~al.
\newblock Noise2music: Text-conditioned music generation with diffusion models.
\newblock {\em arXiv preprint arXiv:2302.03917}, 2023.

\bibitem[\protect\citeauthoryear{Huang \bgroup \em et al.\egroup }{2023b}]{huang2023make}
Rongjie Huang, Jiawei Huang, Dongchao Yang, Yi~Ren, Luping Liu, Mingze Li, Zhenhui Ye, Jinglin Liu, Xiang Yin, and Zhou Zhao.
\newblock Make-an-audio: Text-to-audio generation with prompt-enhanced diffusion models.
\newblock In {\em Proc. Intl. Conf. Machine Learning (ICML)}, 2023.

\bibitem[\protect\citeauthoryear{Jaegle \bgroup \em et al.\egroup }{2022}]{jaegle2021perceiver}
Andrew Jaegle, Sebastian Borgeaud, Jean-Baptiste Alayrac, Carl Doersch, Catalin Ionescu, David Ding, Skanda Koppula, Daniel Zoran, Andrew Brock, Evan Shelhamer, et~al.
\newblock Perceiver io: A general architecture for structured inputs \& outputs.
\newblock In {\em Proc. Intl. Conf. Machine Learning (ICML)}, 2022.

\bibitem[\protect\citeauthoryear{Karras \bgroup \em et al.\egroup }{2022}]{karras2022elucidating}
Tero Karras, Miika Aittala, Timo Aila, and Samuli Laine.
\newblock Elucidating the design space of diffusion-based generative models.
\newblock In {\em Proc. Conf. Neural Information Processing Systems (NeurIPS)}, 2022.

\bibitem[\protect\citeauthoryear{Kong \bgroup \em et al.\egroup }{2020}]{kong2020hifi}
Jungil Kong, Jaehyeon Kim, and Jaekyoung Bae.
\newblock Hifi-gan: Generative adversarial networks for efficient and high fidelity speech synthesis.
\newblock In {\em Proc. Conf. Neural Information Processing Systems (NeurIPS)}, 2020.

\bibitem[\protect\citeauthoryear{Kreuk \bgroup \em et al.\egroup }{2022}]{kreuk2022audiogen}
Felix Kreuk, Gabriel Synnaeve, Adam Polyak, Uriel Singer, Alexandre D{\'e}fossez, Jade Copet, Devi Parikh, Yaniv Taigman, and Yossi Adi.
\newblock Audiogen: Textually guided audio generation.
\newblock In {\em Proc. Intl. Conf. Machine Learning (ICML)}, 2022.

\bibitem[\protect\citeauthoryear{Kumar \bgroup \em et al.\egroup }{2023}]{kumar2023high}
Rithesh Kumar, Prem Seetharaman, Alejandro Luebs, Ishaan Kumar, and Kundan Kumar.
\newblock High-fidelity audio compression with improved rvqgan.
\newblock In {\em Proc. Conf. Neural Information Processing Systems (NeurIPS)}, 2023.

\bibitem[\protect\citeauthoryear{Lee \bgroup \em et al.\egroup }{2023}]{lee2022bigvgan}
Sang-gil Lee, Wei Ping, Boris Ginsburg, Bryan Catanzaro, and Sungroh Yoon.
\newblock Bigvgan: A universal neural vocoder with large-scale training.
\newblock In {\em Proc. Intl. Conf. Machine Learning (ICML)}, 2023.

\bibitem[\protect\citeauthoryear{Li \bgroup \em et al.\egroup }{2023}]{li2023mert}
Yizhi Li, Ruibin Yuan, Ge~Zhang, Yinghao Ma, Xingran Chen, Hanzhi Yin, Chenghua Lin, Anton Ragni, Emmanouil Benetos, Norbert Gyenge, et~al.
\newblock Mert: Acoustic music understanding model with large-scale self-supervised training.
\newblock {\em arXiv preprint arXiv:2306.00107}, 2023.

\bibitem[\protect\citeauthoryear{Liu \bgroup \em et al.\egroup }{2022}]{liu2022diffsinger}
Jinglin Liu, Chengxi Li, Yi~Ren, Feiyang Chen, and Zhou Zhao.
\newblock Diffsinger: Singing voice synthesis via shallow diffusion mechanism.
\newblock In {\em Proc. AAAI Conf. Artif. Intell. (AAAI)}, 2022.

\bibitem[\protect\citeauthoryear{Liu \bgroup \em et al.\egroup }{2023}]{liu2023audioldm}
Haohe Liu, Zehua Chen, Yi~Yuan, Xinhao Mei, Xubo Liu, Danilo Mandic, Wenwu Wang, and Mark~D Plumbley.
\newblock Audioldm: Text-to-audio generation with latent diffusion models.
\newblock In {\em Proc. Intl. Conf. Machine Learning (ICML)}, 2023.

\bibitem[\protect\citeauthoryear{Mariani \bgroup \em et al.\egroup }{2023}]{mariani2023multi}
Giorgio Mariani, Irene Tallini, Emilian Postolache, Michele Mancusi, Luca Cosmo, and Emanuele Rodol{\`a}.
\newblock Multi-source diffusion models for simultaneous music generation and separation.
\newblock {\em arXiv preprint arXiv:2302.02257}, 2023.

\bibitem[\protect\citeauthoryear{Oord \bgroup \em et al.\egroup }{2016}]{oord2016wavenet}
Aaron van~den Oord, Sander Dieleman, Heiga Zen, Karen Simonyan, Oriol Vinyals, Alex Graves, Nal Kalchbrenner, Andrew Senior, and Koray Kavukcuoglu.
\newblock Wavenet: A generative model for raw audio.
\newblock In {\em ISCA Speech Synthesis Workshop}, 2016.

\bibitem[\protect\citeauthoryear{Popov \bgroup \em et al.\egroup }{2021}]{popov2021grad}
Vadim Popov, Ivan Vovk, Vladimir Gogoryan, Tasnima Sadekova, and Mikhail Kudinov.
\newblock Grad-tts: A diffusion probabilistic model for text-to-speech.
\newblock In {\em Proc. Intl. Conf. Machine Learning (ICML)}, pages 8599--8608. PMLR, 2021.

\bibitem[\protect\citeauthoryear{Prenger \bgroup \em et al.\egroup }{2019}]{prenger2019waveglow}
Ryan Prenger, Rafael Valle, and Bryan Catanzaro.
\newblock Waveglow: A flow-based generative network for speech synthesis.
\newblock In {\em Proc. IEEE Intl. Conf. Acoustics, Speech, Signal Process. (ICASSP)}, 2019.

\bibitem[\protect\citeauthoryear{Rafii \bgroup \em et al.\egroup }{2017}]{rafii2017musdb18}
Zafar Rafii, Antoine Liutkus, Fabian-Robert St{\"o}ter, Stylianos~Ioannis Mimilakis, and Rachel Bittner.
\newblock Musdb18-a corpus for music separation.
\newblock 2017.

\bibitem[\protect\citeauthoryear{Ren \bgroup \em et al.\egroup }{2019}]{ren2019fastspeech}
Yi~Ren, Yangjun Ruan, Xu~Tan, Tao Qin, Sheng Zhao, Zhou Zhao, and Tie-Yan Liu.
\newblock Fastspeech: Fast, robust and controllable text to speech.
\newblock In {\em Proc. Conf. Neural Information Processing Systems (NeurIPS)}, volume~32, 2019.

\bibitem[\protect\citeauthoryear{Ren \bgroup \em et al.\egroup }{2020}]{ren2020popmag}
Yi~Ren, Jinzheng He, Xu~Tan, Tao Qin, Zhou Zhao, and Tie-Yan Liu.
\newblock Popmag: Pop music accompaniment generation.
\newblock In {\em Proc. ACM Multimedia (ACM MM)}, 2020.

\bibitem[\protect\citeauthoryear{Ronneberger \bgroup \em et al.\egroup }{2015}]{ronneberger2015u}
Olaf Ronneberger, Philipp Fischer, and Thomas Brox.
\newblock U-net: Convolutional networks for biomedical image segmentation.
\newblock In {\em Proc. Conf. on Medical Image Computing and Computer Assisted Intervention ({MICCAI})}, 2015.

\bibitem[\protect\citeauthoryear{Schneider \bgroup \em et al.\egroup }{2023}]{schneider2023mo}
Flavio Schneider, Zhijing Jin, and Bernhard Sch{\"o}lkopf.
\newblock Mo$\backslash$\^{} usai: Text-to-music generation with long-context latent diffusion.
\newblock {\em arXiv preprint arXiv:2301.11757}, 2023.

\bibitem[\protect\citeauthoryear{Shen \bgroup \em et al.\egroup }{2023}]{shen2023naturalspeech}
Kai Shen, Zeqian Ju, Xu~Tan, Yanqing Liu, Yichong Leng, Lei He, Tao Qin, Sheng Zhao, and Jiang Bian.
\newblock Naturalspeech 2: Latent diffusion models are natural and zero-shot speech and singing synthesizers.
\newblock {\em arXiv preprint arXiv:2304.09116}, 2023.

\bibitem[\protect\citeauthoryear{Simon \bgroup \em et al.\egroup }{2008}]{simon2008mysong}
Ian Simon, Dan Morris, and Sumit Basu.
\newblock Mysong: automatic accompaniment generation for vocal melodies.
\newblock In {\em Proc. Conf. on Human Factors in Computing Systems ({CHI})}, pages 725--734, 2008.

\bibitem[\protect\citeauthoryear{Siuzdak}{2023}]{siuzdak2023vocos}
Hubert Siuzdak.
\newblock Vocos: Closing the gap between time-domain and fourier-based neural vocoders for high-quality audio synthesis.
\newblock {\em arXiv preprint arXiv:2306.00814}, 2023.

\bibitem[\protect\citeauthoryear{Song \bgroup \em et al.\egroup }{2020}]{song2020score}
Yang Song, Jascha Sohl-Dickstein, Diederik~P Kingma, Abhishek Kumar, Stefano Ermon, and Ben Poole.
\newblock Score-based generative modeling through stochastic differential equations.
\newblock In {\em Proc. Intl. Conf. Learning Representations (ICLR)}, 2020.

\bibitem[\protect\citeauthoryear{Wang \bgroup \em et al.\egroup }{2022}]{wang2022songdriver}
Zihao Wang, Kejun Zhang, Yuxing Wang, Chen Zhang, Qihao Liang, Pengfei Yu, Yongsheng Feng, Wenbo Liu, Yikai Wang, Yuntao Bao, et~al.
\newblock Songdriver: Real-time music accompaniment generation without logical latency nor exposure bias.
\newblock In {\em Proc. ACM Multimedia (ACM MM)}, 2022.

\bibitem[\protect\citeauthoryear{Wang \bgroup \em et al.\egroup }{2023}]{wang2023neural}
Chengyi Wang, Sanyuan Chen, Yu~Wu, Ziqiang Zhang, Long Zhou, Shujie Liu, Zhuo Chen, Yanqing Liu, Huaming Wang, Jinyu Li, et~al.
\newblock Neural codec language models are zero-shot text to speech synthesizers.
\newblock {\em arXiv preprint arXiv:2301.02111}, 2023.

\bibitem[\protect\citeauthoryear{Wu \bgroup \em et al.\egroup }{2022}]{wu2022jukedrummer}
Yueh-Kao Wu, Ching-Yu Chiu, and Yi-Hsuan Yang.
\newblock Jukedrummer: conditional beat-aware audio-domain drum accompaniment generation via transformer vq-va.
\newblock In {\em Proc. Intl. Soc. for Music Information Retrieval Conf. ({ISMIR})}, 2022.

\bibitem[\protect\citeauthoryear{Wu \bgroup \em et al.\egroup }{2023}]{wu2023large}
Yusong Wu, Ke~Chen, Tianyu Zhang, Yuchen Hui, Taylor Berg-Kirkpatrick, and Shlomo Dubnov.
\newblock Large-scale contrastive language-audio pretraining with feature fusion and keyword-to-caption augmentation.
\newblock In {\em Proc. IEEE Intl. Conf. Acoustics, Speech, Signal Process. (ICASSP)}, 2023.

\bibitem[\protect\citeauthoryear{Ye \bgroup \em et al.\egroup }{2023}]{ye2023comospeech}
Zhen Ye, Wei Xue, Xu~Tan, Jie Chen, Qifeng Liu, and Yike Guo.
\newblock Comospeech: One-step speech and singing voice synthesis via consistency model.
\newblock In {\em Proc. ACM Multimedia (ACM MM)}, 2023.

\bibitem[\protect\citeauthoryear{Zeghidour \bgroup \em et al.\egroup }{2021}]{zeghidour2021soundstream}
Neil Zeghidour, Alejandro Luebs, Ahmed Omran, Jan Skoglund, and Marco Tagliasacchi.
\newblock Soundstream: An end-to-end neural audio codec.
\newblock {\em IEEE/ACM Trans. Audio, Speech, Lang. Process.}, 30:495--507, 2021.

\end{thebibliography}

\end{document}